\documentclass[english,hyperref]{article}
\usepackage{pslatex}
\usepackage[T1]{fontenc}
\usepackage[latin1]{inputenc}
\usepackage{array}
\usepackage{float}
\usepackage{amsmath}
\usepackage{color}
\usepackage{graphicx}
\usepackage{amssymb}

\makeatletter

\providecommand{\tabularnewline}{\\}

\usepackage{hyperref}

\usepackage{babel}
\makeatother
\begin{document}

\title{Immune System -- Tumor Efficiency Rate as a new Oncological Index
for Radiotherapy Treatment Optimization}

\author{O. Sotolongo-Grau, D. Rodríguez-Pérez, J. A. Santos-Miranda,\\
 O. Sotolongo-Costa, J. C. Antoranz}

\date{~}

\maketitle
\begin{abstract}
A dynamical system model for tumor -- immune system interaction together
with a method to mimic radiation therapy are proposed. A large population
of virtual patients is simulated following an ideal radiation treatment.
A characteristic parameter, the Immune System -- Tumor Efficiency
Rate ($ISTER$), is introduced. $ISTER$ dependence of treatment success
and other features is studied. Statistical results allow us to give
a patient classification scheme. Radiotherapy treatment biological
effective dose ($BED$) is thus optimized based on the patient physical
condition, following the \emph{ALARA} (\emph{As Low As Reasonably
Achievable}) criterion.
\end{abstract}

\section{Introduction}

Some approaches to cancer growth and behavior have been made in the
past years. Recent techniques try to use a population dynamics model
\cite{kuznetzov,sachs,galach,radio1} to mathematically describe the
tumour behavior and its interaction with the immune system. Some of
these works explain tumour behavior under clinical treatments like
cytokines \cite{physD} or radiovirotherapy \cite{rvtherapy} and
properly explain the qualitative behaviors of several tumours. Even
though great efforts had been made to describe cancer radiotherapy
treatments \cite{sachs}, they are but vaguely linked to clinical
observations and their large number of variables and coefficients
make their results hardly transposable to a clinical context.

Radiotherapy and surgery are the most effective treatments for cancer,
and even while surgery has a longer tradition, radiotherapy is replacing
surgery for the control of many tumours \cite{steel}. Those treatments
follow strict protocols that often apply a fixed physical radiation
dose, hardly taking into account the kind of tumour or the patient
immunological condition. Thus, a radiotherapy protocol might result
in a very low success probability for some patients starting their
treatments with a weakened immune system. In practice such a treatment
will be interrupted if the patient physical condition worsens, although
the patient will have already received inappropriate doses of radiation.

Due to its importance and looking for an applicable method, we intend
to model a radiotherapy treatment making the simplest possible assumptions.
Furthermore we will introduced the Immune System Tumour Efficiency
Rate parameter ($ISTER$) as a measure of the patient immune system
strength to fight back cancer. This parameter allow us to make a patient
classification and find the success probability of each patient group
following a radiotherapy treatment protocol. Finally, we will use
these results to assess the optimized biological effective dose ($BED$)
or tissue effect ($E$) based on a given patient physical condition.

\section{Model}

In order to describe the tumour evolution, we propose a Lotka-Volterra
like model based on some assumptions. Tumour cells growth $\dot{X}$
(as usual, a dot over a quantity represents its time derivative) depends
on the current tumour population as $aX$ and its mass-law interaction
with lymphocytes, $-bXY$. Lymphocytes population grows due to tumour-immune
system interaction, $dXY$, and falls in time exponentially, $-fY$,
due to natural cell death. Tumour secretes interleukin which produces
an immune depression effect \cite{depress1,depress2}, and we will
make the simplest assumption supposing it proportional to the tumour
cell number,$-kX$. The tumour is localized and there is a constant
flow, $u$, of lymphocytes from the immune system into this region.

So, we will model tumour-immune system interaction using the known
equations \cite{physD}:

\begin{equation}
\begin{array}{l}
\dot{X}=aX-bXY\\
\dot{Y}=dXY-fY-kX+u\end{array}\label{eq:general}\end{equation}

The effects of radiation over any tissue are generally classified
in three phases \cite{steel}. Physical phase, when radiation ionizes
atoms. Chemical phase, when ionized molecules interact with other
biological components of the cell. And finally, biological phase,
where the damage is fixed, and unrepairable cells are signaled to
die by apoptosis. 

Carcinogenesis and other malignant effects, that escape cellular control,
can appear as late effects of the biological phase and, to avoid them,
radiation doses need to be optimized. This means that higher doses
that could reduce the long term overall survival of patients \cite{survival},
must be avoided whenever possible. 

We will collect all these heterogeneous effects, according to their
time scale, in two groups: short and long term effects. Short term
effects occur at very small time scales compared with the time scales
on which our model runs (those times for which changes in the $X$
and $Y$ variables become appreciable), and so only long term effects
will be taken into account in our evolution equations. Then, we are
going to assume that lymphocytes die or loose their ability to attack
tumour cells immediately, and that radiation dose is concentrated
at an infinitesimal instant of time. At that very moment, long term
effects start to take place, whereas short term effects instantaneously
modify the state of the system.

Thus, we also assume that when a radiation dose is applied at a given
instant $T_{n}$, it induces a fraction $B_{t}$ of the tumour cells
to lose their reproductive endowment and to die exponentially. The
fraction $S_{t}$ of tumour cells not affected by radiation can be
computed by the linear-quadratic (LQ) model \cite{radio1,radio2}, 

\begin{equation}
S_{t}=1-B_{t}=\exp[-E]=\exp[-\alpha\Delta-\beta\Delta^{2}]\label{eq:survt}\end{equation}
where $E$ is known as the tissue effect, $\alpha$ and $\beta$ are
Type A and B damage coefficients \cite{steel}, and $\Delta$ is the
physical radiation dose expressed in Gy, as usual in clinical contexts.
Furthermore, a fraction $B_{l}$ of lymphocytes is also killed by
radiation, in a manner similar to \eqref{eq:survt} although having
different $\alpha$ and $\beta$ coefficients.

To include long term processes in Eqs. \eqref{eq:general}, we write
a new equation for non-proliferating  tumour cells $Z$ \cite{rvtherapy},
taking into account that lymphocyte population is also stimulated,
as $pZY$, due to interaction with these cells. The number of non-proliferating
tumour cells decays exponentially as $-rZ$ due to the death of damaged
cells, and also as $-qZY$ due to the interaction with lymphocytes.
Then we arrive to the system

\begin{equation}
\begin{array}{l}
\dot{X}=aX-bXY-B_{t}(T)X\\
\dot{Y}=dXY+pZY-fY-k(X+Z)+u-B_{l}(T)Y\\
\dot{Z}=B_{t}(T)X-rZ-qZY\end{array}\label{eq:radio}\end{equation}
where $B_{t}(T)=B_{t}\sum\delta(T-T_{n})$ and $B_{l}(T)=B_{l}\sum\delta(T-T_{n})$.
$T_{n}$ are the time instants when radiation doses are applied and
$\delta(T-T_{n})$ denotes Dirac's delta centered at $T_{n}$. We
have supposed that lymphocytes interact in different ways with $X$
and $Z$ cells, although both kind of tumour cells cause the same
depression over the immune system.

Equations \eqref{eq:radio} can be expressed in a dimensionless form
taking the tumour duplication time $\tau_{c}=1/a$ (in absence of
external influences) as the characteristic time, so we introduce the
dimensionless time $\tau=T/\tau_{c}$. Through the substitutions $X=ax/d$,
$Y=ay/b$, $Z=az/d$, we obtain the dimensionless system:

\begin{equation}
\begin{array}{l}
\dot{x}=x-xy-\gamma_{t}(\tau)x\\
\dot{y}=xy+\epsilon zy-\lambda y-\kappa(x+z)+\sigma-\gamma_{l}(\tau)y\\
\dot{z}=\gamma_{t}(\tau)x-\rho z-\eta zy\end{array}\label{eq:aradio}\end{equation}
with $\gamma_{l}=B_{l}/a$, $\gamma_{t}=B_{t}/a$, $\epsilon=p/d$,
$\lambda=f/a$, $\kappa=kb/ad$, $\sigma=ub/a^{2}$, $\rho=ra/d$
and $\eta=qa^{2}/db$.

A linear stability analysis of the system \eqref{eq:aradio} shows
that tumour will vanish to $L_{0}=(0;\sigma/\lambda;0)$ if $\sigma/\lambda>1$
and will remain controlled around $L_{1}=((\lambda-\sigma)/(1-\kappa);1;0)$
if $\kappa<\sigma/\lambda<1$ or $\kappa>1$ \cite{physD}. If the
system is $L_{0}$-stable and initial tumour size is small enough,
then the radiation treatment is unnecessary, whereas if tumour size
is large enough, then the treatment will take it closer to $L_{0}$.

The $L_{1}$ controlled growth state will be reached only if both
parameters fulfill the same condition, in other words, if $\sigma/\lambda$
and $\kappa$ are both greater or smaller than unity at the same time.
Any other condition makes $L_{1}<0$, and even when the stable point
mathematically exists, it can not be approximated from realistic initial
conditions (that should remain positive along the simulation time).
For those patients with $\kappa>1$ and $\sigma/\lambda<1$, the main
effects of the tumour will be the depression of immune system, they
will present a low Karnofsky performance scale \cite{survival} and
will not fulfill physical conditions to be subject under treatment. 

\begin{figure}[h]
\begin{centering}
\includegraphics[scale=0.5]{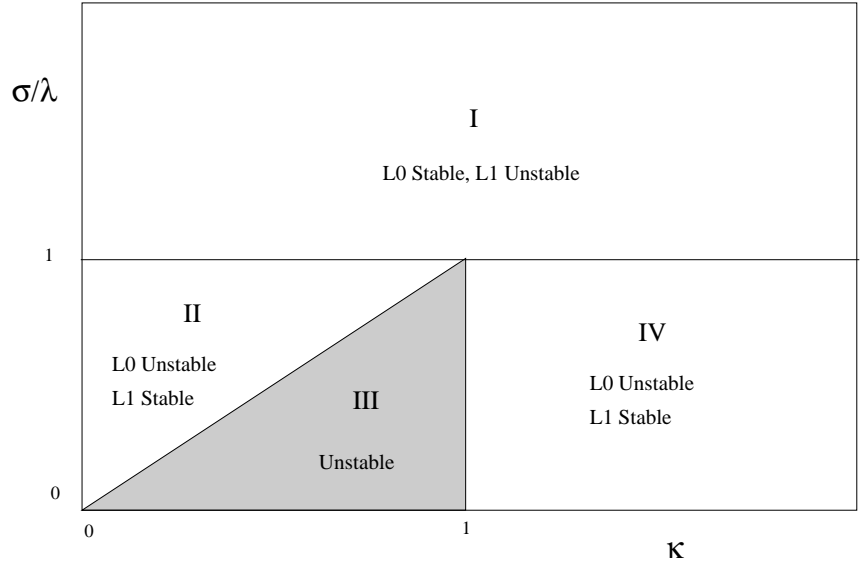}
\par\end{centering}

\caption{\label{fig:stab}Phase diagram of equations \ref{eq:aradio}. The
focus of this work falls inside shadowed region.}
\end{figure}

However for $\sigma/\lambda<\kappa<1$, tumour will grow exponentially
and radiotherapy goal will be to bring it close enough to $L_{0}$
so that immune system can get rid of the tumour. Figure \ref{fig:stab}
shows stable and unstable regions of Eqs. \eqref{eq:aradio} and highlights
region III on which this work will focus. 

The chosen characteristic time and the dimensionless parameters allow
us to give a very intuitive interpretation of the critical parameters
of Eqs. \eqref{eq:aradio}. We can see $\sigma/\lambda$ as the efficiency
of immune system over tumour growth and $\kappa$ as the {}``deficiency''
of the immune system due to tumour growth. 

It is also easy to see that radiation treatments do not change the
stability conditions of our system, given that radiotherapy does not
change tumour or lymphocytes growth rate, but can drive the number
of both kind of cells to very small values. Although Eqs. \eqref{eq:aradio}
allow for infinitesimal $x$ values, in real systems when the number
of tumour cells becomes small enough, immune system may kill them
\cite{steel}. However, in other cases when a few tumour cells survive,
they can cause tumour regrowth. It is known that this behavior is
almost independent on tumour size and as an estimation we will assume
that the closer $L_{0}$ is (in terms of the phase space of figure
\ref{fig:stab}) to the line where it becomes an stable point, the
higher will be the probability of tumour elimination by the immune
system. Thus, when $x$ becomes small enough we will take 

\begin{equation}
P(\sigma/\lambda)=\left\{ \begin{array}{cc}
\sigma/\lambda & \mbox{if}\;\sigma/\lambda<1\\
1 & \mbox{if}\;\sigma/\lambda\geq1\end{array}\right.\label{eq:preg}\end{equation}
as the probability of tumour regression.

Similarly, whenever lymphocyte population becomes zero we will assume
a general failure of immune system. This situation may also occur
in some cases where the tumour is removed but the immune system reaches
such an extreme low concentration of lymphocytes that, consequently,
the patient dies.

\section{Simulation}

We can mimic different radiation treatments with Eqs. \eqref{eq:aradio}
to simulate tumour evolution. To follow radiotherapy treatment in
a realistic way, we apply a radiation session every workday and none
in weekends. All treatments \cite{treat1,treat2} begin the tenth
day, take $6$ weeks of radiotherapy and patients are under observation
until 6 months after the end of radiotherapy sessions. We generate
several virtual patients under treatment taking different values for
the parameter values in Eqs. \ref{eq:aradio} and use a four step
Runge-Kutta method \cite{nr} to integrate them. 

\begin{figure}[h]
\begin{centering}
\includegraphics[scale=0.5]{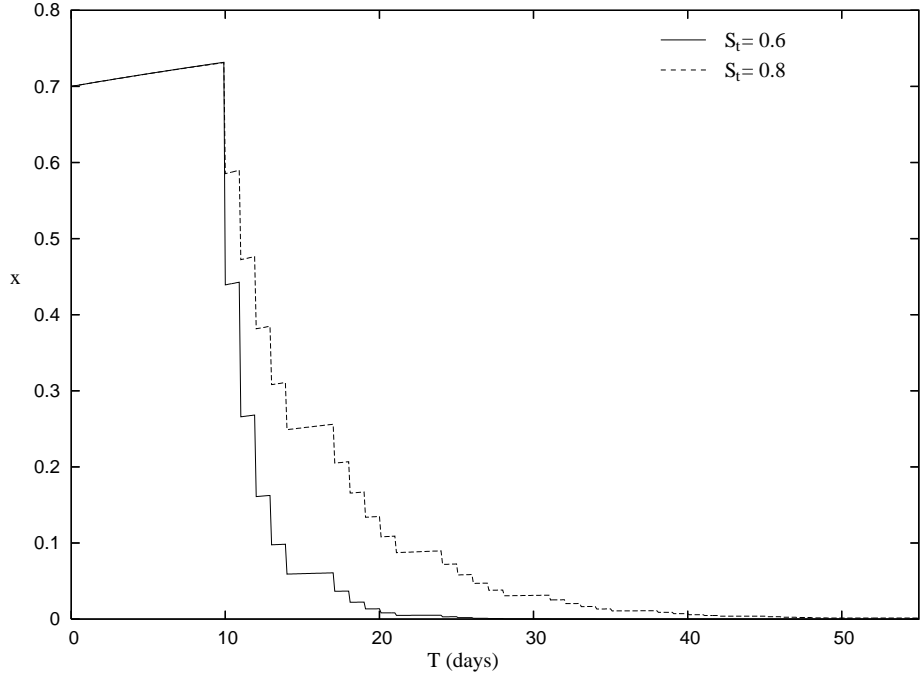}
\par\end{centering}

\caption{\label{fig:tevol}Tumour evolution under radiotherapy treatment for
two different tumour survival factors. $\lambda=5.0$, $\sigma=4.0$,
$\kappa=0.7$, $S_{l}=0.2$}
\end{figure}

To reproduce tumour evolution resembling that of a clinical case,
we need to calculate the correct values of the coefficients appearing
in Eqs. \eqref{eq:general}. Estimation of these coefficients was
made in \cite{retardo}, and following a similar procedure it would
not be too hard for clinical professionals to estimate their values.
Figure \ref{fig:tevol} shows treatment evolution for typical values
of the coefficients and under different doses of radiation. We can
see how the number of tumour cells capable of mitosis quickly decreases
with the radiation therapy. For long enough times, if regression behavior
is not accomplished, the tumour regrows quickly. 

In order to accomplish an statistical study of the dependence of treatment
success on the dosage, and due to the wide range of possible parameter
values in Eqs. \eqref{eq:aradio}, their values are drawn randomly
from a log-normal distribution, to avoid negative values, but keeping
the efficiency of immune system ($\sigma/\lambda$) always smaller
than $1$. Survival factors \cite{radio1,steel} are also taken as
random values within the interval shown in table \ref{tab:EValues}.
As initial conditions we have supposed, for simplicity, that the number
of tumour cells is higher than the number of lymphocytes and that
both populations are distributed as normal random numbers, with parameters
shown in table \ref{tab:SValues}. We have also tested other distributions
for the initial conditions as well as for coefficient values, to verify
that the choice does not affect the qualitative nature of our results. 

\begin{table}[H]
\begin{centering}
\begin{tabular}{|c|c|c|}
\hline 
Parameter&
Minimum&
Maximum\tabularnewline
\hline
\hline 
$\lambda$&
$10^{0}$&
$10^{3}$\tabularnewline
\hline 
$\sigma$&
$10^{-1}$&
$10^{5}$\tabularnewline
\hline 
$\kappa$&
$10^{-2}$&
$10^{4}$\tabularnewline
\hline 
$S_{t}$&
$0.5$&
$0.9$\tabularnewline
\hline 
$S_{l}$&
$0.1$&
$0.4$\tabularnewline
\hline
\end{tabular}
\par\end{centering}

\caption{\label{tab:EValues}Dimensionless parameter values of equations \ref{eq:aradio},
taken from \textcolor{blue}{\cite{retardo,radio1,radio2}}.}
\end{table}

At this point we can proceed to make statistical predictions by generating
a population of {}``virtual patients'' (characterised by their immune
system and tumour parameter values) and simulating their treatment
evolutions. Tables \ref{tab:CValues} and \ref{tab:SValues} show
parameter values used to generate virtual patients.

\begin{table}[H]
\begin{centering}
\begin{tabular}{|c|c|c|}
\hline 
Coefficient&
Mean&
Standard deviation\tabularnewline
\hline
\hline 
$S_{t}$&
0.6&
0.1\tabularnewline
\hline 
$S_{l}$&
0.18&
0.06\tabularnewline
\hline 
$x_{0}$&
1.0&
0.1\tabularnewline
\hline 
$y_{0}$&
0.5&
0.1\tabularnewline
\hline
\end{tabular}
\par\end{centering}

\caption{\label{tab:SValues}Statistical survival factors and initial conditions
for tumour cells and T-lymphocytes}
\end{table}

\begin{table}[H]
\begin{centering}
\begin{tabular}{|c|c|c|}
\hline 
Parameter&
log. Mean&
log. Standard Deviation\tabularnewline
\hline
\hline 
$\lambda$&
5.0&
0.5\tabularnewline
\hline 
$\sigma$&
2.5&
0.5\tabularnewline
\hline 
$\kappa$&
0.8&
0.2\tabularnewline
\hline
\end{tabular}
\par\end{centering}

\caption{\label{tab:CValues}Statistical parameter values.}
\end{table}

\section{Results and clinical interpretation}

\begin{figure}[H]
\begin{centering}
\includegraphics[width=0.5\paperwidth]{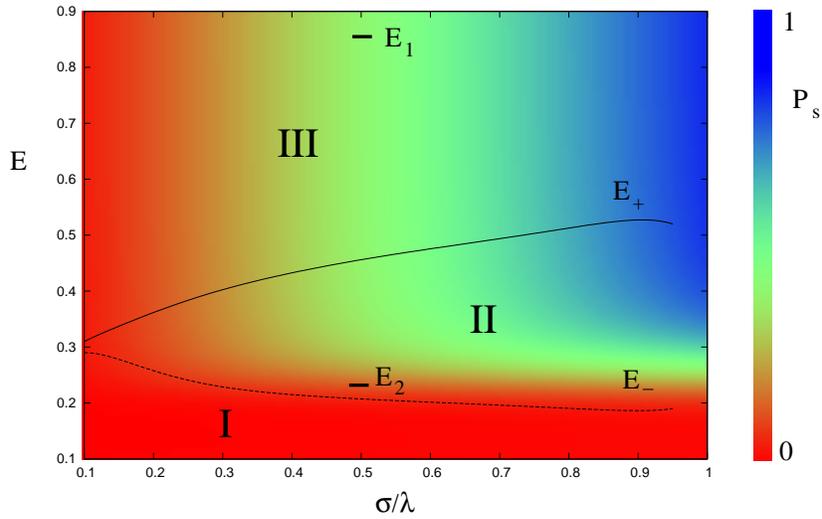}
\par\end{centering}

\caption{\label{fig:Pscolor}$P_{s}$ as a function of $E$ and $\sigma/\lambda$.
$E_{-}$ curve corresponds to the minimum values of tissue effect
for which $P_{s}>0$. $E_{+}$ curve represents the values of tissue
effect for which $P_{s}$ reaches its maximum value, given a fixed
$\sigma/\lambda$. Marks $E_{1}=0.84$ and $E_{2}=0.24$ represents
the tissue effect in the explained examples.}
\end{figure}

We have created a database consisting of over $3\times10^{5}$ virtual
patients. We have calculated the probability of treatment success
($P_{s}$) as the fraction of patients without tumour at the end of
treatment. We have represented this probability $P_{s}$ as a function
of tissue effect, $E$, (see equation \eqref{eq:survt}) and efficiency
of immune system ($\sigma/\lambda$) or $ISTER$. In Fig \ref{fig:Pscolor},
a color map of $P_{s}$ versus $E$ and $\sigma/\lambda$ is represented.
This allows us to classify patients based on their $\sigma/\lambda$--value
and to assess those patients to whom, having an extremely low success
probability, the application of high radiation doses would render
useless. Radiotherapy is not the appropriate treatment for those patients,
although it could be used as a palliative, if a good balance between
drawbacks and advantages is presumed for a specific patient. 

We can see that, for a given value of $\sigma/\lambda$, two significant
values of $E$ can be defined: $E_{-}$, below which $P_{s}$ is very
small (less than $1\%$), and $E_{+}$, above which $P_{s}$ is almost
constant (with less than $1\%$ of change). Results can be fitted
to the expression,\begin{equation}
\begin{array}{c}
P_{s}=P_{s}(\sigma/\lambda,E)\end{array}\label{eq:padj}\end{equation}
and the significant values of $E$ computed as functions of $\sigma/\lambda$.
These two threshold values ($E_{-}$ and $E_{+}$) divide the phase
space $(\sigma/\lambda,E)$ into three regions as shown in Fig \ref{fig:Pscolor}.
The success probability is negligible in region \emph{I}, below $E_{-}$,
while it almost attains its maximum value above $E_{+}$, in region
\emph{III} . However, on the intermediate region \emph{II}, as $E$
grows, $P_{s}$ increases faster towards its maximum value (above
the $E_{+}$ curve). 

The coefficients $\alpha$ and $\beta$ are generally hard to find,
but not impossible, and several values of the ratio $\alpha/\beta$
are reported in the literature \cite{steel}. However this is not
enough for the clinical application and at least one of them must
be found (as explained also in \cite{steel}) to proceed. Luckily,
to characterize patients, we just need, among all coefficients involved
in Eqs. \eqref{eq:radio}, to know the ratio $u/f$, the effective
amount of lymphocytes in the absence of tumour effects or inmunodepression,
and $b/a$, a measure of the effectiveness of lymphocytes over tumour
growth. Clinical professionals must determine the inquiries and tests
needed to find a patient's $ISTER$.

To illustrate a possible clinical application of this result, we are
going to suppose two virtual patients with the same $ISTER=0.5$ and
different tumour sensitivities. We will assume a sensitive tumour
\cite{dale01} with $\alpha=0.3\mbox{ Gy}^{-1}$ and $\alpha/\beta=5\mbox{ Gy}$,
and a more resistant tumour with $\alpha=0.1\mbox{ Gy}^{-1}$ and
$\alpha/\beta=10\mbox{ Gy}$, so the tumour resistance to radiation
is quite different for each case. In both cases the tissue effects,
with an usual treatment, are represented in Fig \ref{fig:Pscolor},
like $E_{1}$ and $E_{2}$ respectively.

First, let us consider the case of a sensitive tumour \cite{dale01}
with $\alpha=0.3\mbox{ Gy}^{-1}$ and $\alpha/\beta=5\mbox{ Gy}$.
If we apply the typical fractionated radiotherapy used in our calculations,
then the biological effective dose ($BED=E/\alpha$), for each radiotherapy
session of $2\mbox{ Gy}$, will be $2.8\mbox{ Gy}$. However,  this
high dose value does not really increase the success probability.
A patient with an immune system efficiency of $ISTER=0.5$, has his
maximum healing probability for a $BED$ value around $1.5\mbox{ Gy}$
in each radiotherapy session. Then, we must apply a physical radiation
of $1.2\mbox{ Gy}$ in each radiation session and thus, avoid an useless
amount of $24\mbox{ Gy}$ to be applied in the whole treatment.

However, in the case of a tumour having a higher resistance to radiation,
e.g. with $\alpha=0.1\mbox{ Gy}^{-1}$ and $\alpha/\beta=10\mbox{ Gy}$,
the same patient with an $ISTER=0.5$, attains a maximum success probability
for a $BED=4.5\mbox{ Gy}$, and needs a physical dose of $3.4\mbox{ Gy}$
to be applied in each session. Besides, the minimum $BED$ value is
$2.2\mbox{ Gy}$, corresponding to $1.85\mbox{ Gy}$ of physical radiation
per session. Table \ref{tab:Optimal-BED} shows the optimal calculated
values of physical radiation dose for the two examples of tumour with
a $ISTER=0.5$.

The oncologist, should decide the amount of radiation to apply, by
evaluating Eq. \eqref{eq:padj} to know the treatment success probability,
and taking into account any other clinical factors implied.

\begin{table}[h]
\begin{tabular}{|>{\centering}p{0.25\columnwidth}|>{\centering}p{0.14\columnwidth}|>{\centering}p{0.14\columnwidth}|>{\centering}p{0.14\columnwidth}|>{\centering}p{0.14\columnwidth}|}
\hline 
Kind of tumour&
$BED$ for $\Delta=2\mbox{ Gy}$&
Optimal $E$&
Optimal $BED$&
Optimal $\Delta$ per session\tabularnewline
\hline
\hline 
\begin{tabular}{c}
Sensitive tumour\tabularnewline
$\alpha=0.3\mbox{ Gy}$\tabularnewline
$\alpha/\beta=5\mbox{ Gy}$\tabularnewline
\end{tabular}&
$2.8\mbox{ Gy}$&
$0.45$&
$1.5\mbox{ Gy}$&
$1.2\mbox{ Gy}$\tabularnewline
\hline 
\begin{tabular}{c}
Less sensitive tumour\tabularnewline
$\alpha=0.1\mbox{ Gy}$\tabularnewline
$\alpha/\beta=10\mbox{ Gy}$\tabularnewline
\end{tabular}&
$2.4\mbox{ Gy}$&
$0.45$&
$4.5\mbox{ Gy}$&
$3.4\mbox{ Gy}$\tabularnewline
\hline
\end{tabular}

\caption{\label{tab:Optimal-BED}Optimal values for two different tumours
with a rate $ISTER=0.5$.}
\end{table}

The presented results match with those reported in \cite{survival},
that show that the long term survivance of patients is not better
at higher doses of radiation. On the contrary, the higher number of
long term survival patients is reached at intermediate doses (between
$2$ or $3\mbox{ Gy}$), even with a smaller total amount of radiation.

\section{Conclusions}

The proposed method, allows us to find the success probability of
a fractionated radiotherapy treatment, using the patient $ISTER$
parameter, as a new oncological index, and the survival fraction $S_{t}$
of tumour cells, even if other parameters involved are unknown. This
calculation provides a way to classify patients, based on their $ISTER$
value, and to approach to the optimum treatment.

The radiotherapy treatment must be designed for each patient taking
into account his/her immunological characteristics ($ISTER$) relative
to the tumour. Tissue effect has to be tuned to be larger than $E_{-}$,
otherwise no success will be achieved, but needs not to be larger
than $E_{+}$, because no improvement will be obtained for larger
radiation doses. Thus, in accordance with the ALARA (\emph{As Low
As Reasonably Achievable}) principle \cite{alara}, the physical radiation
doses should be adjusted to bring $E$ as close as possible to $E_{+}$
but without out-ranging it. This optimization process could be performed
once the clinical professionals find a way to evaluate the $ISTER$
index experimentally for a given patient. On other hand, the values
of of $\alpha$ and $\beta$ (in Eq. \ref{eq:survt}) are known or
feasible to find for many kinds of tumour.

\appendix

\section{Fitting result data to an analytical function}

We used a Levenberg-Marquardt \cite{nr} method to fit the result
data, showed in figure \ref{fig:Pscolor}, to the analytical function,\begin{equation}
\frac{P_{s}(\sigma/\lambda,E)}{\sigma/\lambda}=\left(\theta+\phi\left(1+\left(\frac{E+\psi}{\varphi}\right)^{4}\right)^{-1}\right)\label{eq:ps_simple}\end{equation}
for each value of computed $\sigma/\lambda$. This expression gives
us a family of functions related to each other through coefficients
$\theta$, $\phi$, $\psi$ and $\varphi$. These coefficients are
functions of only $\sigma/\alpha$ and can be easily fitted using
the same numerical method. 

We have found the following numerical expressions for these coefficients,

\begin{equation}
\begin{array}{c}
\theta=0.950271*\left(1-exp(-4.66627\frac{\sigma}{\lambda}-0.24319)\right)\\
\phi=-0.935012+exp(-4.71719\frac{\sigma}{\lambda}-0.289458)\\
\varphi=0.0450091\frac{\sigma}{\lambda}+0.091267\\
\psi=0.0159581\frac{\sigma}{\lambda}-0.141425\end{array}\label{eq:ps_coefs}\end{equation}

Merging all this expressions, it is possible analyze the behaviour
of the success probability $P_{s}(\sigma/\lambda,E)$.
\end{document}